\documentclass[final,twoside,twocolumn]{siamltex}
\usepackage{amsmath}
\usepackage{amsfonts}
\usepackage{graphics}
\usepackage{epsfig}
\def\norm#1{\left\|#1\right\|}
\newcommand{\abs}[1]{\lvert#1\rvert}

\newcommand{\sen}{\mathop{\rm sen}\nolimits}

\title{The method of moments for  Nonlinear Schr\"{o}dinger Equations:
Theory and Applications.}

\author{ V\'{\i}ctor M. P\'erez-Garc\'{\i}a
    \thanks{Universidad de Castilla-La Mancha, Departamento de Matem\'aticas,
    E.T.S.I. Industriales, Avd. Camilo Jos\'e Cela, s/n, Ciudad Real, E-13071,
    Spain. ({\tt victor.perezgarcia@uclm.es}), supported by grants BFM2003-02832 (Ministerio de Educaci\'on y Ciencia) and PAI-05-001
    (Consejer\'{\i}a de Educaci\'on y Ciencia de la Junta de Comunidades de Castilla-La Mancha).}
     \and
     Pedro Torres \thanks{Universidad de Granada, Departamento
de Matem\'aticas, Campus de Fuentenueva s/n, Granada 18071, supported  by D.G.I. BFM2002-01308,
Ministerio Ciencia y Tecnolog\'{\i}a, Spain.}
   \and
   Gaspar D. Montesinos
   \thanks{Universidad de Castilla-La Mancha,
Departamento de Matem\'aticas,
    E.T.S.I. Industriales, Avd. Camilo Jos\'e Cela, s/n, Ciudad Real, E-13071,
    Spain, supported by grants BFM2003-02832 (Ministerio de Educaci\'on y Ciencia) and PAI-05-001
    (Consejer\'{\i}a de Educaci\'on y Ciencia de la Junta de Comunidades de Castilla-La Mancha).}
    }

\begin{document}

\maketitle

\begin{abstract}
The method of moments in the context of Nonlinear Schr\"odinger Equations relies on defining a set of integral quantities, which characterize the solution of this partial differential equation and whose evolution can be obtained from a set of ordinary differential equations. In this paper we find all cases in which the method of moments leads to closed evolution equations, thus extending and unifying previous works in the field of applications. For some cases in which the method fails to provide rigorous information we also develop approximate methods based on it, which allow to get some approximate  information on the dynamics of the solutions of the Nonlinear Schr\"odinger equation.
\end{abstract}



\pagestyle{myheadings} \thispagestyle{plain} \markboth{V. M.
P\'EREZ-GARC\'IA, P. TORRES, G. D. MONTESINOS}{THE METHOD OF MOMENTS FOR
NLS EQUATIONS}


\section{Introduction}

Nonlinear Schr\"{o}dinger (NLS) equations  appear in a great array of
contexts \cite{Vazquez} as for example in
semiconductor electronics \cite{Soler,Soler2}, optics in nonlinear
media \cite{Kivshar}, photonics \cite{Hasegawa}, plasmas
\cite{Dodd}, fundamentation of quantum mechanics
\cite{fundamentals}, dynamics of accelerators \cite{Fedele}, 
mean-field theory of Bose-Einstein condensates \cite{Dalfovo} or 
in  biomolecule dynamics \cite{Davidov}. In some of these fields and many others, the NLS equation
appears as an asymptotic limit for a slowly varying dispersive
wave envelope propagating in a nonlinear medium \cite{scott}.

The study of these equations has served as the catalyzer of the
development of new ideas or even mathematical concepts such as solitons \cite{Zakharov}
or singularities in EDPs \cite{Sulembook,SIAMFibich}.

One of the most general ways to express a nonlinear
Schr\"{o}dinger equation,  is
\begin{equation}
\label{NLS} i\frac{\partial u }{\partial t} =
-\frac{1}{2}\Delta u + V(x,t)u + g(|u |^{2},t)u -i\sigma (t)u
\end{equation}
where $\Delta = \partial^2/\partial x_1^2 + \cdots +
\partial^2/\partial x_n^2$ and $u$ is a complex function which describes some physical wave.
We shall consider here the solution of \eqref{NLS} on
$\mathbb{R}^n$ and therefore $u: \mathbb{R}^n \times [0,T]
\rightarrow \mathbb{C}$, with initial values $u(x,0) = u_0(x) \in X$, $X$ being an appropriate
functional space ensuring finiteness of certain integral quantities to be defined later.

The family of nonlinear Schr\"{o}dinger equations \eqref{NLS}
contains many particular cases, depending on the specific choices of the nonlinear terms
$g(|u|^2,t)$, the potentials $V(x,t)$, the dissipation $\sigma(t)$ and the dimension of the space $n$. The
best known cases are those of power-type 
$g(|u|^{2}) =  \alpha |u|^p$ or polinomial  $g(|u|^{2})= \alpha_1 |u|^{p_1} + \alpha_2 |u|^{p_2}$
nonlinearities.

The potential term $V(x,t)$ models the action of an external force
acting upon the system and may have many forms. Finally, we include in Eq. (\ref{NLS}) a simple loss term
arising in different applications \cite{Perez95,Fibich2001}. In many cases these losses are negligible, i. e. $\sigma =0$.

The description of the dynamics of initial data ruled by this
equation is of great interest for applications. Nevertheless,
gathering information about the solutions of a partial
differential equation that is nonlinear like \eqref{NLS}
constitutes a problem that is {\em a priori} nearly
unapproachable. For this reason, most studies about the dynamics of this type of
equation are exclusively numerical. The rigorous studies carried
out to date concentrate on (i) properties of stationary solutions
\cite{stationary}, (ii) particular results on the existence of
solutions \cite{Vaz,DCDSB}, and (iii) asymptotic properties
\cite{Sulembook,SIAMFibich}. 

Only when $n=1, g(\abs{u}^2,t) = \abs{u}^2, V(x,t) = 0, \sigma=0$ it is possible to arrive at a solution of the initial value problem
by using the inverse scattering transform method \cite{Zakharov}. 
In this paper we develop the so-called method of moments, which tries to provide qualitative information about
the behavior of the solutions of nonlinear Schr\"{o}dinger
equations. Instead of tackling the Cauchy problem
(\ref{NLS}) directly, the method studies the evolution of several
 integral quantities (the so-called \emph {moments}) of the
solution $u(x,t)$. In some cases the method allows to reduce
the problem to the study of systems of coupled ordinary nonlinear
differential equations. In other cases the method provides a
foundation for making approximations in a more systematic (and 
simpler) way than other procedures used in physics, such as those involving finite-dimensional reductions of the original problem: namely the averaged Lagrangian, collective coordinates or variational methods \cite{Borisreview,Angel}.
 In any case the method of moments
provides an information which is very useful for the applied
scientist, who is usually interested in obtaining as much
information as possible characterizing the \emph{dynamics}
 of the solutions of the problem.

It seems that the first application of the method of moments was
performed by Talanov \cite{Talanov} in order to find a formal
condition of sufficiency for the blowup of solutions of the
nonlinear Schr\"{o}dinger equation with $g(|u|^2)=-|u|^2$ and
$n=2$. Since then, the method has been applied to different
particular cases (mainly solutions of radial symmetry in two spatial dimensions),
especially in the context of optics where many equations of NLS type arise \cite{Porras}.

In previous researchs, the method of moments has been studied in a range of
specific situations but in all the cases the success of the method
is unrelated to a more general study. In this paper we try to
consider the method systematically and solve a number of open
questions: (i) to find the most general type of nonlinear term and potentials in Eq. \eqref{NLS} for
which the method of moments allows to get conclusions and (ii) to develop approximate
methods based on it for situations in which the moment equations do not allow
 to obtain exact results.

\section{Preliminary considerations}

Let us define the functional space $Q(H)$ as the space of functions for
which the so called energy functional
\begin{equation}\label{energia4}
E(u) = (u,Hu)_{L^2(\mathbb{R}^n)} + \int_{\mathbb{R}^n} G(|u|^2,t) dx
\end{equation}
is finite, $G$ being a function such that $\partial
G(|u|^2,t)/\partial |u|^2 = g(|u|^2,t)$,
$(\cdot,\cdot)_{L^2(\mathbb{R}^n)}$ denoting the usual scalar
product in $L^2(\mathbb{R}^n)$ and $H = -\frac{1}{2}\Delta +
V(x,t)$.
For the case $V(x,t) = 0$ and $g$ independent of time, several
results on existence and unicity were given by Ginibre
and Velo \cite{Vaz}.

As regards the case $V \neq 0$, which is the one we are mainly
interested in our work, the best documented case in the literature
is that of potentials with $|D^{\alpha}V|$ bounded in
$\mathbb{R}^n$ for every $|\alpha|\geq 2$; that is, potentials
with at most quadratic growth. Y. Oh \cite{Oh} proved the local existence of solutions
in $L^2(\mathbb{R}^n)$ and in $Q(H)$ for nonlinearities of the
type $g(u) = -|u|^{p}, 0 \leq p < 4/n$. However the procedure
used allows to substitute this nonlinearity with other more
general ones. It seems also posible to extend the results
to the case in which the potential depends on $t$.

Therefore, from now on we shall suppose that the nonlinear term
satisfies the conditions set by Ginibre and Velo and
that  $|D^{\alpha}V|$ is bounded in space for all  $|\alpha|\geq
2$. Under these conditions it is posible to guarantee at least local
existence of solutions of Eq. \eqref{NLS} in appropriate functional spaces.

\subsection{Formal elimination of the loss term}

\label{perdido} In the first place we carry out the
transformation \cite{Perez95}
\begin{equation}
\hat{u}(x,t) = u(x,t)e^{\int_{0}^t\sigma(\tau)d\tau},
\end{equation}
which is well defined for any bounded function $\sigma(t)$ (this includes all known realistic cases arising in the applications). The equation satisfied by $\hat{u}(x,t)$ is
obtained from the following direct calculation:
\begin{eqnarray*}
\label{NLSp} i\frac{\partial \hat{u} }{\partial t} & = & i
\frac{\partial u}{\partial t} e^{\int_{0}^t\sigma(\tau)d\tau}+ i
\sigma(t) u
e^{\int_{0}^t\sigma(\tau)d\tau} \\
& = & \left[-\frac{1}{2}\Delta u + V(x,t)u + g(|u |^{2},t)u
-i\sigma (t)u\right]e^{\int_{0}^t\sigma(\tau)d\tau} + i
\sigma(t) u e^{\int_{0}^t\sigma(\tau)d\tau}\\
& = & -\frac{1}{2}\Delta \hat{u} + V(x,t)\hat{u} +
\hat{g}(|\hat{u}|^{2},t)\hat{u}
\end{eqnarray*}
where
\begin{equation}\hat{g}(|\hat{u}|^{2},t) = g(|u|^2,t) =
g(e^{-2\int_{0}^t\sigma(\tau)d\tau}|\hat{u}|^2,t).
\end{equation}

 From
here on we shall consider, with no loss of generality, that
$\sigma(t)=0$ in Eq. (\ref{NLS}) assuming that this choice might add an extra time-dependence to the nonlinear term.

\section{The method of moments: Generalities}
\label{moment}

\subsection{Definition of the moments}

Let us define the following quantities:
\begin{subequations}
\label{momenta}
\begin{eqnarray}
I_{k,j}(t) & = & \int x_j^k |u(x,t)|^2 dx = \norm{x_j^{k/2}u}^2_{L^2(\mathbb{R}^n)}, \\
V_{k,j}(t) & = & 2^{k-1} i \int x_j^k \left(u(x,t) \frac{\partial
\bar{u(x,t)}}{\partial x_j} -
 \bar{u} \frac{\partial u(x,t)}{\partial x_j}\right) dx,\\
K_j(t) & = & \frac{1}{2}\int \left|\frac{\partial u(x,t)}{\partial
x_j}\right|^2 dx = \frac{1}{2}\norm{\frac{\partial u}{\partial
x_j}}^2_{L^2(\mathbb{R}^n)} \\
J(t)  & = & \int G(|u(x,t)|^2,t) dx,\label{momentJ}
\end{eqnarray}
\end{subequations}
with $j=1,...,n$ and $k = 0, 1, 2, ...$, which we will call
\emph{moments} of $u(x,t)$ in analogy with the moments of a
distribution. From now on and also in Eqs. (\ref{momenta}) it is understood that all integrals and norms refer to the spatial variables $x \in \mathbb{R}^n$ unless otherwise stated. In Eqs. (\ref{momenta}) we denote by $\bar{u}$
the complex conjugate of $u$.

In some cases we will make specific reference to which solution
$u$ of Eq. \eqref{NLS} is used to calculate the moments by means
of the notation: $I^{u}_{k,j}$, etc.

The moments are quantities that have to do with \emph{intuitive}
properties of the solution $u(x,t)$. For example, the moment
$I_{0,0}$ is the squared $L^2(\mathbb{R}^n)$-norm of the solution
and therefore measures the \emph{magnitude}, \emph{quantity} or
\emph{mass} thereof. Depending on the particular context of
application, this moment is denominated mass, charge, intensity,
energy, number of particles, etc. The moments $I_{1,j}(t)$ are the
coordinates of the \emph{center} of the distribution $u$, giving
us an idea of the overall position thereof. The quantities
$I_{2,j}$ are related with the \emph{width} of the distribution
defined as $W_j = (\int_{\mathbb{R}^n} (x_j-I_{1,j})^2 |u|^2 dx)^{1/2} =
(I_{2,j}+I_{1,j}^2 I_{0,0} - 2 I_{1,j}^2)^{1/2}$, which is also a quantity with an evident
meaning. 


The evolution of the moments is determined by that of the function
$u(x,t)$. From now on we will assume that the initial datum $u_0(x)$ and the properties of the equation guarantee that the moments are well defined for all time.

\subsection{First \emph{conservation law}}

It is easy to prove formally that the moment
$I_{0,0}$ is invariant during the temporal evolution by just
calculating
\begin{eqnarray}
\frac{d}{dt} I_{0,0}(t) & = & \int_{\mathbb{R}^n}  \left(\frac{d}{dt} |u|^2\right)
dx = \int_{\mathbb{R}^n} \left(\bar{u} \frac{\partial}{\partial t}u + u
\frac{\partial}{\partial t}\bar{u} \right) dx \nonumber \\
& = & \int_{\mathbb{R}^n}  i\left(\frac{1}{2}\bar{u} \Delta u -
V(x,t)|u|^2 - g(|u|^{2},t)|u|^2\right) dx \nonumber \\
& & + \int_{\mathbb{R}^n}  i\left(-\frac{1}{2}u \Delta \bar{u} +
V(x,t)|u|^2 + g(|u|^{2},t)|u|^2\right) dx \nonumber \\
& = & \int_{\mathbb{R}^n}  \frac{i}{2} \left(\bar{u} \Delta u - u \Delta
\bar{u} \right) dx  \nonumber\\ \label{dNdt} & = & \frac{i}{2}
\left( \int_{\mathbb{R}^n}  |\nabla u|^2 dx  - \int_{\mathbb{R}^n}  |\nabla u|^2 dx
 \right) = 0,
\end{eqnarray}
where we have performed integration by parts and used that the function $u$ and its derivatives vanish at infinity.

Obviously the above \emph{demonstration} is formal in the sense
that a regularity which we do not know for certain has been used
for $u$. Nevertheless, this type of proof can be formalized by making a convolution of the function $u$ with a regularizing function. The
details of these methodologies can be seen in \cite{Vaz} or
\cite{Oh}. In this paper we will limit ourselves to formal
calculations.



\section{General results for harmonic potentials}
\label{casoarmonico}

\subsection{Introduction}

From this point onward we will focus on the particular case of
interest for this study when $V(x,t)$ is a harmonic potential of
the type $V(x,t) = \frac{1}{2}(x,\Lambda(t) x)$, where $\Lambda$ is a real matrix of
the form $\Lambda_{ij}(t) = \lambda_{i}^2(t) \delta_{ij}$, with
$\lambda_i \geq 0$ for $i=1,...,n$ and $\delta_{ij}$ is the Kronecker delta. Bearing in mind the results of
Section \ref{perdido}, the NLS equation under study is then
\begin{eqnarray}
\label{NLSredu} i\frac{\partial u }{\partial t} =
-\frac{1}{2}\Delta u + \frac{1}{2}\left(\sum_{j=1}^n
\lambda_j^2 x_j^2 \right) u + g(|u |^{2},t)u.
\end{eqnarray}
This equation appears in a wide variety of applications such as
propagation of waves optical transmission lines with online
modulators \cite{Smith,Gabitov,Kumar,Turitsyn}, propagation of
light beams in nonlinear media with a gradient of the refraction
index \cite{us1,us2}, or dynamics of Bose-Einstein condensates
\cite{Dalfovo}. Generically it can provide a model for studying
some properties of the solutions of nonlinear Schr\"{o}dinger equations localized near a
minimum of a general potential $V(x)$.

\subsection{First moment equations}

If we differentiate the definitions of the moments $I_{1,j}$ and
$V_{0,j}$, we obtain, after some calculations, the evolution
equations
\begin{subequations}
\begin{eqnarray}
\frac{dI_{1,j}}{dt} & = & V_{0,j},\\
\frac{dV_{0,j}}{dt} & = & -\lambda_j^2 I_{1,j},
\end{eqnarray}
\end{subequations}
so that $I_{1,j}, j=1,...,n$ satisfy
\begin{equation}\label{43}
\frac{d^2I_{1,j}}{dt^2} + \lambda_j^2 I_{1,j} = 0
\end{equation}
with inital data $I_{1,j}(0), \dot{I}_{1,j}(0) = V_{0,j}(0)$.
These expressions are a generalization of the
Ehrenfest theorem of linear quantum mechanics to the nonlinear Schr\"odinger equation
and  particularized for
the potential that concerns us \cite{Oh,Perez98}.

This result has been discussed previously in many papers and is
physically very interesting. It indicates that the evolution of
the center of the solution is independent of the nonlinear effects
and of the evolution of the rest of the moments and depends only on the potential parameters.

\subsection{Reduction of the general problem to the case $\boldsymbol{I_{1,j} = V_{0,j} = 0}$}

We shall begin by stating the following lemma \cite{Vadym}:
\begin{lemma}
Let $u(x,t)$ be a solution of \eqref{NLSredu} with the initial
datum $u(x,0) = u_0(x)$. Then, the functions
\begin{subequations}
\begin{equation}
\label{tra1}
 u_{R}(x,t) = u(x-R(t),t) e^{i\theta(x,t)}
\end{equation}
where
\begin{equation}
\label{tra2} \theta(x,t) = \left(x,\dot{R}\right) + \int_0^t
\left[(\dot{R}(t'),\dot{R}(t'))-(R(t'),\Lambda(t') R(t'))\right]
dt'
\end{equation}
and
\begin{equation}
\label{tra3} \frac{d^2R}{dt^2} + \Lambda R = 0
\end{equation}
for any set of initial data $R(0), \dot{R}(0) \in \mathbb{R}^n$ are also solutions of \eqref{NLSredu}.
\end{subequations}
\end{lemma}
\begin{proof} All we have to do is substitute
\eqref{tra1}, \eqref{tra2} and \eqref{tra3} in \eqref{NLSredu}.
\end{proof}

One noteworthy conclusion is that given a solution of Eq. \eqref{NLSredu} we can
\emph{translate} it initially by a  constant vector and obtain another
solution. In the case of stationary states, defined as solutions 
of the form 
\begin{equation}\label{st}
 u(x,t) = \varphi_{\mu}(x) e^{i\mu t}
 \end{equation}
  which exist in the autonomous case (i.e. $d\lambda/dt=0$) and 
whose dynamics are
trivial, this result implies that under displacements the only
dynamics acquired are those of the movement of the center given by Eq. \eqref{tra3}. The coincidence of the evolution laws \eqref{43} and \eqref{tra3}
 allows us to state the following theorem which is an
immediate consequence of the above lemma.

\begin{theorem} If $\psi(x,t)$ is a solution of \eqref{NLSredu}  with
non-zero $I^{\psi}_{1,j}$ or $V^{\psi}_{0,j}$, then there exists a
unique solution $u(x,t) =
\psi(x+\{I^{\psi}_{1,j}(t)\}_j,t)e^{i\theta(x,t)}$ with $$
\theta(x,t) =-\sum_j x_j V^{\psi}_{0,j} + \sum_j \left[\int_0^t
V^{\psi}_{0,j}(t')^2 - \lambda_j(t')^2 I_{1,j}^{\psi}(t')^2
\right] dt'$$ such that $I^{u}_{1,j} = 0$ and $V^{u}_{0,j} = 0$
\label{VadVad}
\end{theorem}

The important conclusion of this theorem is that it suffices to
study solutions with $I_{1,j}$ and $V_{0,j}$ equal to zero, as those
that have one of these coefficients different from zero can be
obtained from previous ones, by means of translation and
multiplication by a linear phase in $x$. From a practical
standpoint, what is most important is that $I_{1,j}$ be null
without any loss of generality, as then we can establish a direct
link between the widths and the moments $I_{2,j}$ (see the
discussion in the third paragraph of Section \ref{moment}).

\subsection{Moment Equations}

Assuming that all of the moments can be defined at any time $t$, we can
calculate their evolution equations by means of direct
differentiation. The results are gathered in the next theorem.

\begin{theorem} Let $u_0(x)$ be an initial datum such that the moments $I_{2,j}, V_{1,j}, K_j$ and $J$ are well defined at $t=0$. Then
\begin{subequations}
\label{momentos-teor}
\begin{eqnarray}
\frac{dI_{2,j}}{dt} & = & V_{1,j},\label{widtha}\\
\frac{dV_{1,j}}{dt} & = & 4K_j-2\lambda_j^2I_{2,j}-2\int_{\mathbb{R}^n} D(\rho,t) dx,\label{widthb}\\
\frac{dK_j}{dt} & = & -\frac{1}{2}\lambda_j^2V_{1,j}-\int_{\mathbb{R}^n} D(\rho,t)
\frac{\partial ^2\phi}{\partial x_j^2}\, dx, \label{widthc} \\
\frac{dJ}{dt} & = & \int_{\mathbb{R}^n} D(\rho,t) \Delta\phi\, dx + \int_{\mathbb{R}^n}
\frac{\partial G(\rho,t)}{\partial t} dx\label{widthd}.
\end{eqnarray}
\end{subequations}
where $D(\rho,t) = G(\rho,t)- \rho g(\rho,t)$, $\rho = |u(x,t)|^2$.
\end{theorem}
\begin{proof}
The demonstration of the validity of Eqs. \eqref{momentos-teor} can be
carried out from direct calculations, performing integration by
parts, and using the decay of $u$ and $\nabla u$ at infinity.

To demonstrate \eqref{widtha} it is easier to work with the
modulus-phase representation of $u$, $u = \rho^{1/2} e^{i\phi}$ (with $\rho>0$).
Then
\begin{eqnarray*}
\frac{dI_{2,j}}{dt} & = & \int x_j^2 \dot{\rho} =  - \int x_j^2
 (\nabla \rho\cdot\nabla \phi + \rho \Delta\cdot \phi) \\
& = & - \int x_j^2  \nabla \rho\cdot \nabla \phi + \int \nabla (
x_j^2\rho) \cdot \nabla \phi \\ & = & - \int x_j^2  \nabla
\rho\cdot \nabla \phi + \int x_j^2 \nabla \rho\cdot\nabla \phi + \int
\nabla ( x_j^2) \rho\cdot\nabla \phi \\ & = & 2 \int x_j \rho
\frac{\partial \phi}{\partial x_j} = V_{1,j}.
\end{eqnarray*}

We can also prove \eqref{widthb} as follows
\begin{eqnarray}
\frac{dV_{1,j}}{dt} & = & i \int x_j \left(u_t
\frac{\partial\bar{u}}{\partial x_j}+u
\frac{\partial\bar{u_t}}{\partial x_j}-\bar{u_t} \frac{\partial
u}{\partial x_j}-\bar{u}\frac{\partial u_t}{\partial x_j}\right)
\nonumber\\
 & = & \int x_j\left[-\frac{1}{2}\Delta
u\frac{\partial\bar{u}}{\partial
x_j}+\frac{1}{2}\left(\sum\lambda_k^2
x_k^2\right)u\frac{\partial\bar{u}}{\partial x_j}+g u
\frac{\partial\bar{u}}{\partial x_j} + u
\frac{1}{2}\frac{\partial\Delta\bar{u}}{\partial
 x_j} \right. \nonumber \\
 &  & \left. -\lambda_j^2 x_j|u|^2-\frac{1}{2}\left(\sum\lambda_k^2
x_k^2\right)u\frac{\partial\bar{u}}{\partial x_j}-\frac{\partial
g}{\partial x_j}|u|^2-g u\frac{\partial\bar{u}}{\partial
x_j}+\mbox{c.c.}\right] \nonumber\\
 & = & -2\lambda_j^2\int x_j^2|u|^2 - 2 \int x_j|u|^2\frac{\partial g}{\partial x_j} 
 \nonumber\\
 &  & -\frac{1}{2}\int x_j\left(\Delta u\frac{\partial\bar{u}}{\partial
 x_j}+\Delta\bar{u}\frac{\partial u}{\partial x_j}-u\frac{\partial\Delta\bar{u}}{\partial
 x_j}+\bar{u}\frac{\partial\Delta u}{\partial x_j}\right),
 \label{V1j}\\ \nonumber
\end{eqnarray}
where c.c. indicates the complex conjugate. Operating on the above
integrals we have
\begin{eqnarray*}
\lefteqn{\int x_j\left(\Delta u\frac{\partial\bar{u}}{\partial
 x_j}+\Delta\bar{u}\frac{\partial u}{\partial x_j}-u\frac{\partial\Delta\bar{u}}{\partial
 x_j}-\bar{u}\frac{\partial\Delta u}{\partial x_j}\right) } \\
& & = -2\int x_j \left(\nabla
 u\cdot\frac{\partial\nabla\bar{u}}{\partial
 x_j}+\nabla\bar{u}\cdot\frac{\partial\nabla u}{\partial x_j}\right)
 - 4\int\left|\frac{\partial u}{\partial
 x_j}\right|^2-2\int|\nabla u|^2 \\ & & = 2\int|\nabla u|^2 - 4\int\left|\frac{\partial u}{\partial
 x_j}\right|^2-2\int|\nabla u|^2 = -4\int\left|\frac{\partial u}{\partial
 x_j}\right|^2,
\end{eqnarray*}
and
\begin{equation*}
\int x_j\frac{\partial g}{\partial x_j}\rho =  -\int g\rho-\int
x_j\frac{\partial G}{\partial x_j} = -\int g\rho+\int G = \int D.
\end{equation*}
By substitution in \eqref{V1j} we arrive at the desired result.
\begin{eqnarray*}
\frac{dV_{1,j}}{dt} & = & -2\lambda_j^2\int
x_j^2|u|^2+2\int\left|\frac{\partial u}{\partial
 x_j}\right|^2-2\int D\\
 & = & -2\lambda_j^2 I_{2,j}+4 K_j-2\int D.
\end{eqnarray*}

Let us now prove \eqref{widthc}
\begin{eqnarray*}
\frac{dK_j}{dt} & = & \frac{1}{2} \int \frac{d}{dt}
\left(\frac{\partial u}{\partial
x_j}\frac{\partial\bar{u}}{\partial x_j}\right)  =  \frac{1}{2}
\int\left(\frac{\partial u_t}{\partial
x_j}\frac{\partial\bar{u}}{\partial x_j} + \frac{\partial
u}{\partial x_j}\frac{\partial\bar{u_t}}{\partial x_j}\right) \\
 & = & \frac{1}{2} \int\frac{\partial}{\partial
 x_j}\left[\frac{i}{2}\Delta u-\frac{i}{2}\left(\sum_{k=1}^n \lambda_k^2
 x_k^2\right) u-igu\right]\frac{\partial\bar{u}}{\partial
 x_j}+\mbox{c.c.},\\
\end{eqnarray*}
then
\begin{eqnarray*}
 \frac{dK_j}{dt} & = & -\frac{i}{2}\lambda_j^2\int x_j \left(u \frac{\partial
\bar{u}}{\partial x_j} -
 \bar{u} \frac{\partial u}{\partial x_j}\right)-\frac{i}{2}\int\frac{\partial g}{\partial x_j}\left(u
\frac{\partial \bar{u}}{\partial x_j} -
 \bar{u} \frac{\partial u}{\partial
 x_j}\right)\\
 &  & +\frac{i}{4}\int\left(\frac{\partial\Delta
 u}{\partial x_j}\frac{\partial\bar{u}}{\partial x_j}-
 \frac{\partial u}{\partial x_j}\frac{\partial\Delta\bar{u}}{\partial
 x_j}\right) =  -\frac{1}{2}\lambda_j^2 V_{1,j}+\int
 g\frac{\partial}{\partial
 x_j}\left(\rho\frac{\partial\phi}{\partial
 x_j}\right)
\\
 & = & -\frac{1}{2}\lambda_j^2 V_{1,j}+\int
 g\left(\rho\frac{\partial^2\phi}{\partial
 x_j^2}+\frac{\partial\rho}{\partial
 x_j}\frac{\partial\phi}{\partial x_j}\right),\\
\end{eqnarray*}
and using the definition of $G$ we obtain
\begin{eqnarray*}
 \frac{dK_j}{dt} & = & -\frac{1}{2}\lambda_j^2 V_{1,j}+\int
 g\rho\frac{\partial^2\phi}{\partial
 x_j^2}+\int\frac{\partial G}{\partial
 x_j}\frac{\partial\phi}{\partial x_j}\\
 & = &-\frac{1}{2}\lambda_j^2 V_{1,j}+\int
 g\rho\frac{\partial^2\phi}{\partial
 x_j^2}-\int G\frac{\partial^2\phi}{\partial x_j^2}\\
 & = & -\frac{1}{2}\lambda_j^2 V_{1,j}-\int(G-\rho
 g)\frac{\partial^2\phi}{\partial x_j^2} =
 -\frac{1}{2}\lambda_j^2 V_{1,j}-\int
D\frac{\partial^2\phi}{\partial x_j^2}.
\end{eqnarray*}

Finally, to demonstrate \eqref{widthd} we proceed
\begin{eqnarray*}
\frac{dJ}{dt}  & = & \int \frac{dG(\rho,t)}{dt} = \int
\left[\frac{\partial G}{\partial \rho} \left(\frac{\partial
\rho}{\partial u} u_t + \frac{\partial \rho}{\partial \bar{u}}
\bar{u}_t\right) + \frac{\partial G}{\partial t}\right] \\ & = &
\int \left[ \frac{\partial G}{\partial \rho}\left(\bar{u} u_t + u
\bar{u}_t\right) + \frac{\partial G}{\partial t}\right] =
\frac{i}{2} \int \left[g \left( \bar{u} \Delta u - u \Delta
\bar{u}\right) + \frac{\partial G}{\partial t}\right] \\ & = &
\int \left[ - g \nabla\cdot (\rho \nabla \phi) +
\frac{\partial G}{\partial t}\right] =  \int \left[ - g
\nabla \rho\cdot \nabla \phi - g \rho \Delta \phi +
\frac{\partial G}{\partial t}\right]\\ & = & \int \left[ - \nabla
G \cdot \nabla \phi - g \rho \Delta \phi + \frac{\partial
G}{\partial t}\right] = \int \left( G - g \rho\right) \Delta
\phi +  \int \frac{\partial G}{\partial t}\\
 & = & \int D \Delta
\phi +  \int \frac{\partial G}{\partial t}.
\end{eqnarray*}

\end{proof}

A direct consequence of the theorem is
\begin{corollary} \label{estato} Let $u(x,t)$ be a stationary solution of
\eqref{NLSredu}. Then,
\begin{equation}
K_j = \frac{1}{2}\lambda_j^2 I_{2,j} + \frac{1}{2}\int D(\rho) dx.
\end{equation}
\end{corollary}

\section{Solvable cases of the method of moments}
\label{resoluble}

In this section we will study several
particular situations of practical relevance in which the
method of moments thoroughly provides exact  results.

\subsection{The linear case $\boldsymbol{g(\rho,t) = 0}$}

In this case, Eqs. \eqref{momentJ} and (\ref{widthd}) tells us that $J(t) =0$, for all $t$, and then the moment equations \eqref{momentos-teor} become
\begin{subequations}
\label{momentoslineal}
\begin{eqnarray}
\frac{dI_{2,j}}{dt} & = & V_{1,j},\label{linealwa}\\
\frac{dV_{1,j}}{dt} & = & 4K_j-2\lambda_j^2I_{2,j},\\
\frac{dK_j}{dt} & = & -\frac{1}{2}\lambda_j^2V_{1,j}.
\label{linealwc}
\end{eqnarray}
\end{subequations}
That is, in the linear case the equations for the moments along each direction $j$
of the physical  space $\mathbb{R}^n$ are uncoupled. This property was  known
in the context of optics for $n=2$ and constant $\lambda_j$
\cite{Porras2}. Here we see that this property holds for any number of spatial dimensions, time dependence $\lambda(t)$  and 
even for nonsymmetric initial data.

\subsection{Condition of closure of the moment equations in the general case}

Equations \eqref{momentos-teor} do not form a closed set and
therefore to obtain, in general, their evolution we would need to
continue obtaining moments of a higher order, which would
provide us with an infinite hierarchy of differential equations.
Given the similarity among the terms that involve second
derivatives of the phase of the solution in Eqs. \eqref{momentos-teor}, it is natural to wonder whether it would
be possible to somehow close the system and thus obtain
information about the solutions.

From this point on, and for the rest of the section, we will limit
ourselves to the case $\lambda_j(t) = \lambda(t),j=1, \ldots,n$,
which is the most realistic one, and which includes as a
particular case the situation without external potentials $\lambda_j = 0$. Let us define the following
quantities
\begin{equation}
\label{valoresmedios}
\mathcal{I} =  \sum_{j=1}^n I_{2,j}, \
\mathcal{V}  =  \sum_{j=1}^n V_{1,j}, \
\mathcal{K}  =  \sum_{j=1}^n K_{j}.
\end{equation}
Differentiating equations \eqref{valoresmedios} and using
\eqref{momentos-teor}, we have
\begin{subequations}
\begin{eqnarray}
\frac{d\mathcal{I}}{dt} & = & \mathcal{V},\label{wa}\\
\frac{d\mathcal{V}}{dt} & = & 4\mathcal{K}-2\lambda^2\mathcal{I}-2n\int_{\mathbb{R}^n} D(\rho,t) dx, \label{wb} \\
\frac{d\mathcal{K}}{dt} & = & -\frac{1}{2}\lambda^2\mathcal{V}
-\int_{\mathbb{R}^n} D(\rho,t) \Delta \phi dx, \label{wc} \\
\frac{dJ}{dt} & = &  \int_{\mathbb{R}^n} D(\rho,t) \Delta\phi + \int_{\mathbb{R}^n} \frac{\partial
G(\rho,t)}{\partial t} dx\label{wd}.
\end{eqnarray}
\end{subequations}
If we add up equations \eqref{wc} and \eqref{wd} we arrive at
\begin{subequations}
\label{cierre}
\begin{eqnarray}
\frac{d\mathcal{I}}{dt} & = & \mathcal{V},\label{wwa}\\
\frac{d\mathcal{V}}{dt} & = & 4\left[\mathcal{K}-\frac{n}{2}\int_{\mathbb{R}^n}
D(\rho,t) dx\right]-
2\lambda^2\mathcal{I}, \label{wwb}\\
\frac{d\left(\mathcal{K}+J\right)}{dt} & = &
-\frac{1}{2}\lambda^2\mathcal{V} + \int_{\mathbb{R}^n} \frac{\partial G}{\partial
t}dx \label{wwc}.
\end{eqnarray}
\end{subequations}
In order that equations \eqref{cierre} form a closed system, they
must fulfill that $-\tfrac{n}{2}\int_{\mathbb{R}^n} D(\rho,t) dx = J = \int_{\mathbb{R}^n} G(\rho,t) dx$
and that $\int_{\mathbb{R}^n} \frac{\partial G(\rho,t)}{\partial t} dx$ can be expressed in
terms of the other known quantities. The former condition requires
that $$ 0 = \int_{\mathbb{R}^n} \left[ \frac{n}{2} D(\rho,t) + G(\rho,t)\right] dx =
\int_{\mathbb{R}^n} \left[ \left( 1 + \frac{n}{2}\right) G(\rho,t) -
\frac{n}{2}\frac{\partial G(\rho,t}{\partial \rho} \rho \right] dx.$$ As
$G(\rho,t)$ does not depend explicitly on $x$, this condition is
verified when
\begin{equation}
 \frac{\partial G(\rho,t)}{\partial \rho} = \left(1 + \frac{2}{n}\right) \frac{G(\rho,t)}{\rho},
 \end{equation}
 that is, if
 \begin{equation}
G(\rho,t) = g_0(t) \rho^{1+2/n},
 \end{equation}
or, equivalently, if
 \begin{equation}
 g(\rho,t)  = g_0(t) \rho^{2/n},
 \end{equation}
where $g_0(t)$ is an arbitrary function that indicates the
temporal variation of the nonlinear term. Then
\begin{equation}
\int_{\mathbb{R}^n} \frac{\partial
G(\rho,t)}{\partial t} dx = \frac{1}{g_0}\frac{dg_0}{dt} \int_{\mathbb{R}^n} G(\rho,t) dx=
\frac{1}{g_0}\frac{dg_0}{dt} J(t).
\end{equation}
To close the equations it is necessary that $g_0(t)$ be
constant in order to cancel the last term of this expression.
Then,  the nonlinearities for which it is possible to find closed
results are
\begin{equation}\label{ecu-nlin}
g(\rho) = g_0 \rho^{2/n} = g_0 |u|^{4/n},
\end{equation}
with $g_0\in \mathbb{R}$, remembering that in the case $g_0<0$
there may be problems of blowup in finite time. Fortunately,
these nonlinearities for $n=1,2,3$ correspond to cases of
practical interest. For instance, the case $n=1$  with quintic nonlinearity
has been studied in Refs. 
\cite{BEC5a,BEC5b,Christiansen} and the case $n=2$, with cubic
nonlinearity, corresponds probably to  the most relevant instance of NLS equation, i.e. the cubic 
one in two spatial dimensions \cite{us1,us2}. For $n=3$ the nonlinearity given by Eq. \eqref{ecu-nlin} appears 
in the context of the Hartree-Fock theory of atoms.

\subsection{Simplification of the moment equations}

Defining a new quantity $\mathcal{E} =\mathcal{K}+J$, Eqs. \eqref{cierre} become
\begin{subequations}
\label{linealmomentos}
\begin{eqnarray}
\frac{d\mathcal{I}}{dt} & = & \mathcal{V},\label{wwwa}\\
\frac{d\mathcal{V}}{dt} & = & 4\mathcal{E}-2\lambda^2(t) \mathcal{I}, \label{wwwb}\\
\frac{d\mathcal{E}}{dt} & = & -\frac{1}{2}\lambda^2(t) \mathcal{V}
\label{wwwc}.
\end{eqnarray}
\end{subequations}
These equations form a set of non-autonomous linear equations for
the three averaged moments: $\mathcal{E}(t), \mathcal{V}(t)$
and $\mathcal{I}(t)$. To continue our analysis, we note that
\begin{equation}
\mathcal{Q} = 2\mathcal{E}\mathcal{I} - \frac{1}{4}\mathcal{V}^2,
\end{equation}
is a dynamical invariant of Eqs. \eqref{linealmomentos}. We finally define $X =
\left|\mathcal{Q}\right|^{-1/4}\mathcal{I}^{1/2}$, which is proportional to the
\emph{mean width} of $u$. A simple calculation allows us to
corroborate that the equation that $X(t)$ satisfies is
\begin{equation}\label{Pinney}
\frac{d^2X}{dt^2} + \lambda^2(t) X = \frac{\text{sgn}(\mathcal{Q})}{X^3}.
\end{equation}
Solving \eqref{Pinney} allows to calculate $\mathcal{V}$ and
$\mathcal{E}$ by simple substitution in (\ref{linealmomentos}).
This equation is similar to that obtained in \cite{us1} for
solutions of radial symmetry in the case $n=2$ and $g(u) = |u|^2$. Here we find that it is possible to obtain a more
general result for solutions without specific symmetry requirements, and for any combination of dimension and nonlinearity $g(u)=|u|^p$ satisfying the condition
$np=4$. The case with $\text{sgn}(\mathcal{Q}) = -1$ corresponds to collapsing situations \cite{Sulembook,Weinstein}. In what follows we consider mostly the case $\mathcal{Q} > 0$.

Equation (\ref{Pinney}) was studied by Ermakov in 1880
\cite{Ermakov}, although since then it has been \emph{rediscovered} 
many times (see e.g. \cite{Pinney}). It is a particular
case of the so-called Ermakov systems \cite{Reida,tor1,tor2}, for
which it is possible to give fairly complete results. Especially
easy, though tedious to demonstrate is
\begin{theorem}[Ermakov, 1880]
Let $X(t)$ be the solution of \eqref{Pinney} with initial data
$X(0) = X_0, \dot{X}(0) = \dot{X}_0$. Then, if $\chi_1(t)$ and
$\chi_2(t)$ are solutions of the differential equation
\begin{subequations}
\begin{equation}\label{Hill}
\frac{d^2\chi}{dt^2} + \lambda^2(t) \chi = 0
\end{equation}
satisfying the initial data $\chi_1(0) = X_0, \dot{\chi}_1(0) =
\dot{X}_0$ and $\chi_2(0) = 0, \dot{\chi}_2(0) \neq 0$ then
\begin{equation}\label{superposicion}
X(t) = \sqrt{\chi^2_1(t) + \frac{1}{w^2}\chi_2^2(t)}
\end{equation}
\end{subequations}
where $w$ is the constant $w = \chi_1 \dot{\chi}_2 - \chi_2
\dot{\chi}_1$.
\end{theorem}

Equation \eqref{superposicion} is often called \emph{the principle
of nonlinear superposition}. Equation \eqref{Hill} is the
well known Hill's equation \cite{HILL} which modelizes a
parametrically forced oscillator, and which has been studied in
depth. In the following, we shall study a couple of special
situations in view of their physical interest.

It is remarkable that the complex dynamics of a family of
nonlinear partial differential equations can be understood in
terms of a simple equation such as Hill's.

If we suppose that the function $\lambda^2(t)$ depends on a
parameter $\varepsilon$ in the way $\lambda^2(t)=1 + \tilde
\lambda_{\varepsilon}(t)$, $\tilde \lambda_{\varepsilon}(t)$ being
a periodic function with maximum value $\varepsilon$ (not
necessarily small), there exists a complete theory that describes
the intervals of values of $\varepsilon$ for which the solutions
of Eq. (\ref{Hill}) are bounded (intervals of stability) and
the intervals for which the solutions are unbounded (intervals of
instability) \cite{HILL}.

\section{Applications}
\label{aplicaciones}

\subsection{Dynamics of laser beams in GRIN media}
\label{optica}

When a laser beam propagates in a medium with a gradex refraction index (GRIN medium)  with an
specific profile quadratic in the transverse coordinates, 
the distribution of intensity $u(x,y,z)$ in permanent regime is
ruled by Eq. \eqref{NLSredu} with $g(\rho) =
 \rho$ and $n = 2$ (in the optical version of the equation $t
 \leftrightarrow z$), so that we are dealing with the
critical case that we know how to solve. Although in principle it
would be possible to design fibers with arbitrary profiles,
technically the simplest way is to join fibers with different
uniform indexes in each section.

In this case, the modelization of the phenomenon is given by
\begin{equation}
\label{pt} \lambda^2(t)=\begin{cases}
                      a^2 & t \in [0,T_a]\\
                      b^2 & t \in (T_a,T_a + T_b=T]
                      \end{cases}
\end{equation}


Equation (\ref{Hill}) with $\lambda(t)$ given by \eqref{pt} is
known as the Meissner equation whose solution is trivial, 
given in each segment by
a combination of trigonometric functions.

\begin{figure}[tbp]
\begin{center}
\epsfig{file=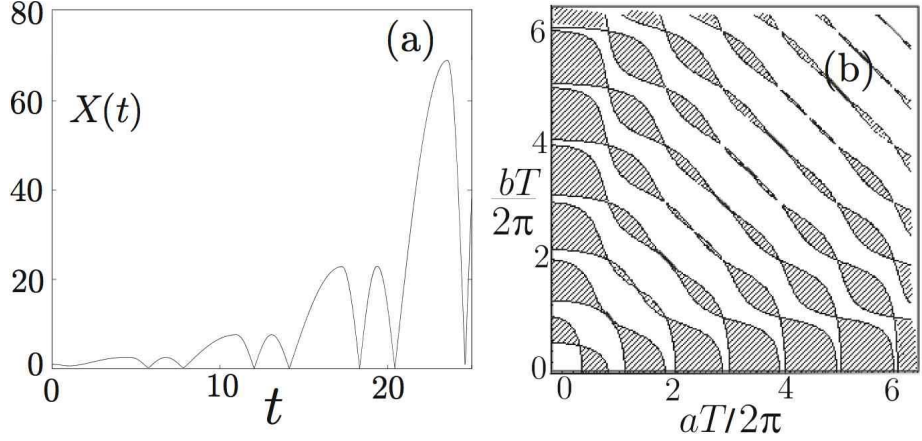,width=0.85\textwidth}\end{center}
\caption{Solutions of equation (\ref{Hill}) with $\lambda(t)$
given by (\ref{pt}). (a) Resonant solution for $T_a = 10\pi, T =
20\pi, a=0.05, b=0.15$ and (b) Regions of
resonance for $T_a=T_b=T/2$.} \label{plotres}
\end{figure}

The solutions to the Meissner equation can be bounded (periodical
or quasiperiodical) or unbounded (resonant oscillations). In
figure \ref{plotres} the two types of solutions are shown for a
particular choice of parameters.

As far as the regions of stability in the space of parameters are
concerned, they can be obtained by studying the discriminant of
equation (\ref{Hill}), defined as the trace of the monodromy
matrix, that is
\begin{equation}
D(a,b,T_a,T_b):=\phi_1(T)+\phi_2'(T).
\end{equation}
where $\phi_1,\phi_2$ are the solutions of (\ref{Hill}) satisfying
the initial data $\phi_1(0)=1,\phi_1'(0)=0$ and
$\phi_2(0)=0,\phi_2'(0)=1$ respectively.

In our case it is easy to arrive at
\begin{subequations}
\begin{eqnarray}
\phi_1(T) & = & \cos(aT_a)\cos(bT_b)
-\frac{a}{b}\sen(aT_a)\sen(bT_b), \\ \phi_2'(T) & = &
\cos(aT_a)\cos(bT_b)-\frac{b}{a}\sen(aT_a)\sen(bT_b).
\end{eqnarray}
\end{subequations}

\begin{figure}
\begin{center}
\epsfig{file=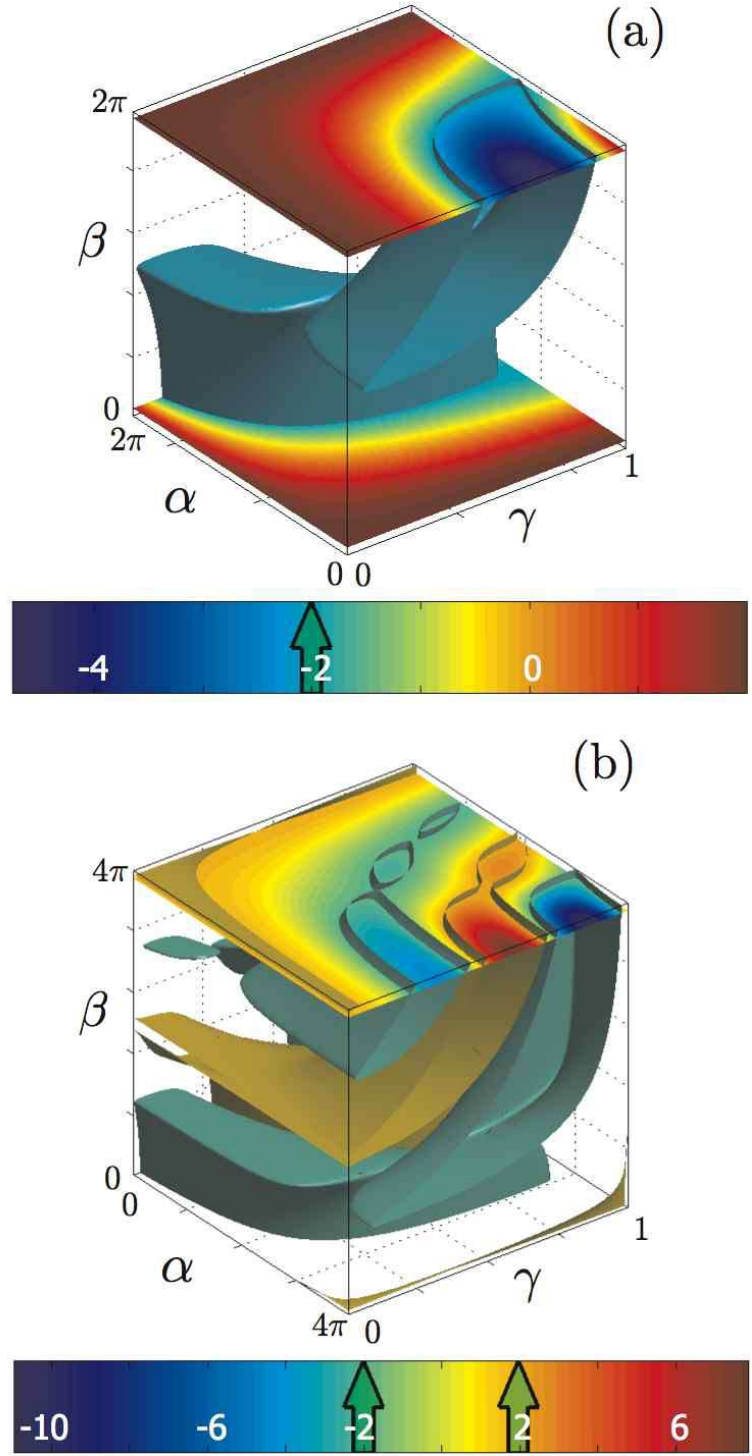,width=0.7\textwidth}
\end{center}
\caption{(a) Isosurfaces corresponding to $D = -2$ (in this range
of values $D\leq 2$) for a limited range of parameters. The
regions between the gray surface and the planes limiting the
drawing are regions of resonance. Sections are shown for two
particular values of $\beta$ where the bluish tones correspond to
the regions of resonance. The color bar indicates the color
corresponding to each level of $D(\gamma,\alpha,\beta)$ and the
arrow indicates the color assigned to the isosurface $D=-2$. (b)
The same as (a) but for a larger range of parameters. Isosurfaces
$D = 2$ and $D=-2$ are shown in brown and green respectively. A
section is shown for a particular value of $\beta$ with bluish and
reddish tones corresponding to regions of resonances with $D<-2$
and $D>2$, respectively. In this case the two values $D = 2$ and
$D=-2$ are indicated by arrows on the color bar. \label{guay}}
\end{figure}

Finally, the form of the discriminant is
\begin{equation}
D(a,b,T_a,T_b) = 2\cos(aT_a+bT_b)- \label{pipito}
\frac{(a-b)^2}{ab}\sen(aT_a)\sen(bT_b)
\end{equation}
The Floquet theory for linear equations with periodical
coefficients connects the stability of the solutions of
(\ref{Hill}) with the value of the discriminant. The regions of
resonance correspond to values of the parameters for which
$|D|>2$, whereas if $|D|<2$ the solutions are bounded
\cite{HILL}. The equations $D(a,b,T_a,T_b)=2$ and
$D(a,b,T_a,T_b)=-2$ are the manifolds that limit the regions of
stability in the 4-dimensional space of parameters. In reality,
defining $\alpha = aT, \beta = bT, T_a = \gamma T, T_b =
(1-\gamma)T$ the number of parameters is reduced to three
\begin{equation}
D(\gamma,\alpha,\beta) = 2\cos(\alpha \gamma + \beta (1-\gamma)) -
\frac{(\alpha-\beta)^2}{\alpha \beta}\sen(\alpha \gamma)\sen(\beta
(1-\gamma)),
\end{equation}
Therefore, the isosurfaces $D(\gamma,\alpha,\beta) = 2$ and
$D(\gamma,\alpha,\beta)=-2$ can be visualized in three dimensions
as is shown in figure \ref{guay}.

The general study of the regions that appear in figure \ref{guay}
is complex, which leads us to focus below on a few particular
cases.

For example, in the case in which the two sections have the same
length $T_a=T_b$, the discriminant depends only on $aT,bT$ and it
is
\begin{equation}
D(a,b) = 2\cos\left(\frac{(a+b)T}{2}\right) -
\frac{(aT-bT)^2}{abT^2}\sen \left(\frac{aT}{2}\right) \sen
\left(\frac{bT}{2}\right)
\end{equation}
so that now the condition $|D| = 2$ determines curves such as
those of figure \ref{plotres}(b).

\begin{figure}[tbp]
\begin{center}
\epsfig{file=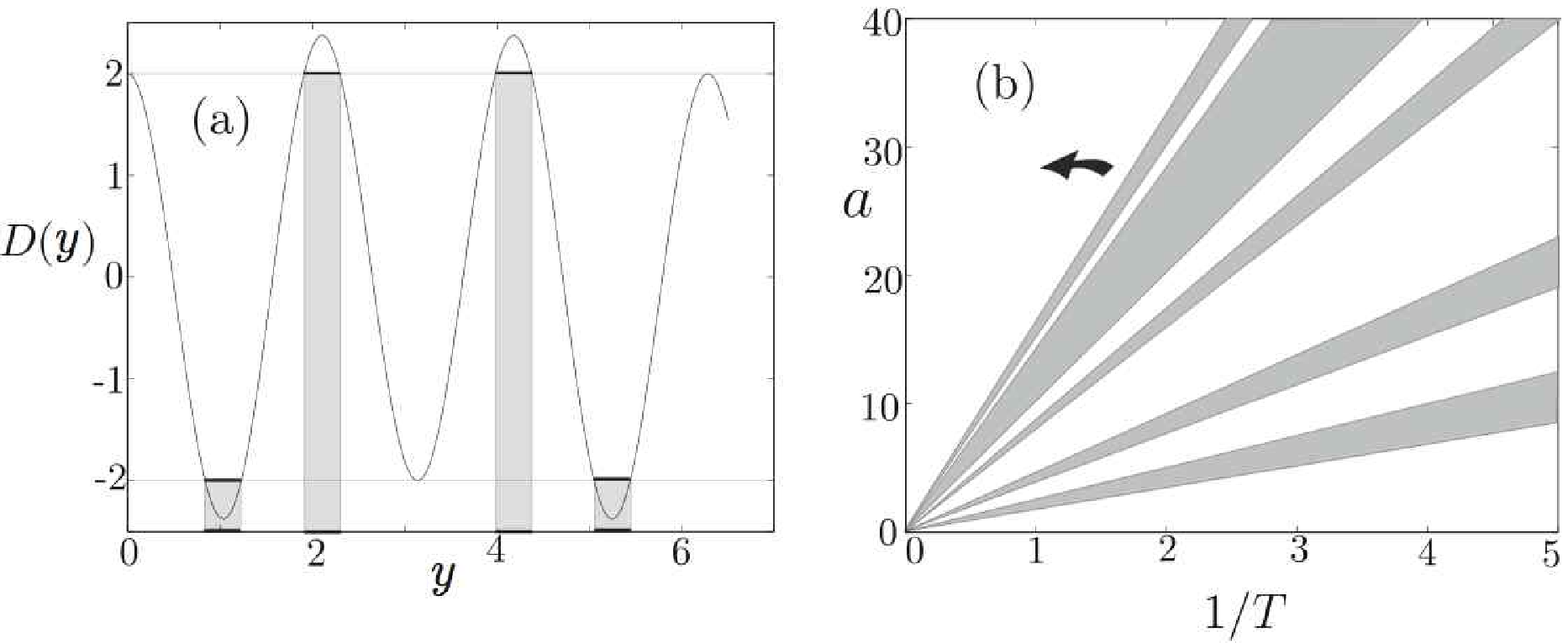,width=0.95\textwidth} \caption{(a)
First four regions of resonance in $y$ (in color) for $T_a = T/2,
b = 2a$, as a function of $y=aT/2$. (b) First five regions of
resonance in the plane $a - 1/T$.}\label{tertio}
\end{center}
\end{figure}

The structure of the regions of resonance can be explored in
more detail fixing the relative values of the coefficients, for
example taking $b = 2a$.
\begin{equation}
D(a,T)=2\cos\left(\frac{3aT}{2}\right)-\frac{1}{2}\sen\left(\frac{aT}{2}\right)\sen(aT)
\end{equation}
Defining the variable $y = aT/2$, the discriminant is a function
$D(y)$ [Fig. \ref{tertio}]. The so-called characteristic curves
are hyperboles of the form $2y_{\pm}^{(n)} = aT$ being
$y_{+}^{(n)}, y_{-}^{(n)}$ respectively the solutions of the
algebraic equations $
f_{\pm}(y) =   2 \cos 3 y - \tfrac{1}{2} \sen y \sen 2y \mp 2 = 0.$
It is easy to demonstrate that the regions of resonance are
contained between two consecutive zeros of $f_+$ or $f_-$ that can
be obtained using any elementary numerical method. If we draw $a$
as a function of $1/T$, the regions of resonance are those marked
in colors in figure \ref{tertio}(b). Obviously the image is
repeated due to the periodicity 2$\pi$ of $D(y)$ and there are
only four basic regions of resonance (together with its harmonics)
contained in the intervals (roots of $f_+$ and $f_-$): $y \in
[0.84,1.23] \cup [1.91,2.3] \cup [3.98,4.37] \cup [5.05,5.44]$
(see figure \ref{tertio}(a)).

Another case of possible interest is that in which one of the
fibers is not of GRIN type, that is, $b = 0$. Then the discriminant
is given by the limit of equation (\ref{pipito}) when
$b\rightarrow 0$.
\begin{equation} \label{discriminantito} D(a,T)=-
aT \sen{\left(\frac{aT}{2}\right)} +
2\cos{\left(\frac{aT}{2}\right)}.
\end{equation}
As in the previous case the only relevant parameter is $y = aT/2$,
the regions of resonance on the plane $a-T$ are hyperboles and the
relevant quantities are the zeroes of $f_+$ and $f_-$, which are
given by
\begin{equation}
\label{zerable3}
f_{\pm}(x) = - y \sen y + 2 \cos y \mp 2 = 0.
\end{equation}
Now there is no exact periodicity in the positions of the zeros
any more, but at least it is possible to estimate the location of
those of high order. To do this we must bear in mind that for $y$
big enough the dominating term in both cases is  $f_{\pm}(y)
\simeq - y \sen y$, so that the zeros will be given by $y = n
\pi$. It can be seen with a perturbative argument that the
convergence ratio is of the order of $\mathcal{O}(1/n)$. Writing
$y_{\pm}^{(n)} = n \pi + \varepsilon^{(n)}_{\pm}$ and substituting
it in equations (\ref{zerable3}) it is found that
\begin{equation}
\label{zerable4} \varepsilon_{\pm}^{(n)} \simeq (-1)^{n+1} \frac{1\pm
2}{n \pi}.
\end{equation}
This type of analysis can be extended to any restricted set of
parameters.

\subsection{Dynamics of Bose-Einstein condensates}

In the case of Bose-Einstein condensates, there has recently been
great interest in the study of the dynamics of these systems in a
parametrically oscillating potential. Recent experiments
 (see e. g. \cite{exp1,exp2}) have motivated a
series of qualitative theoretical analyses (the pioneer works on
this subject can be seen in \cite{teor1,teor2,teor3} although
there is a great deal of posterior literature).

In the models to which we refer, the trap is modified harmonically
in time, that is
\begin{equation}\label{mathieu}
\lambda^2(t) = 1 + \varepsilon \cos \omega t
\end{equation}
with $\varepsilon > -1$. Equation (\ref{Hill}) with $\lambda(t)$
given by \eqref{mathieu} is called the Mathieu equation.
For this equation it is possible, as in the case of the Meissner equation, to carry out the
study of the regions of the space of parameters in which resonances
occur. In the first place, for any fixed $\varepsilon$, there
exist two successions $\{\omega_n\},\, \{\omega_n'\}$ with
$\omega_n,\omega'_n
 \stackrel{n\rightarrow \infty}{\longrightarrow} 0$ such that if
 we take $\omega \in (\omega_n,\omega_n')$,
equation (\ref{Hill}) possesses a resonance. In the second place,
for fixed $\omega$, the resonances appear when $\varepsilon$ is large enough. 
The
boundaries of those regions are the so-called characteristic
curves that cannot be obtained explicitly but whose existence can
be demonstrated analytically, as in the previous section, by using
the discriminant. In the case of the Mathieu equation, it can be
proven that the regions of instability begin in frequencies
$\omega=2,1,1/2,\ldots,2/n^2,\ldots$ \cite{HILL}.

As in the previous case, the resonant behavior depends only on the
parameters, and not on the initial data. With respect to
stability, the Massera theorem implies that if
$(\varepsilon,\omega)$ is in a region of stability, then there exists
a periodic solution of (\ref{Hill}), and by the nonlinear
superposition principle such a solution is stable in the sense of
Liapunov.


\section{Approximate methods (I): Quadratic phase approximation (QPA)}
\label{approx}

\subsection{Introduction and justification of the QPA}

Up to now, the results we have shown for the evolution of the solution moments are exact and in some sense rigorous. Unfortunatelly, in many situations of practical interest it is not possible to obtain closed evolution equations for the moments. In this section we
will deal with an approximate method which is based on the method of moments.

The idea of this method is to approximate the phase of the
solution $u$ by a quadratic function of the coordinates, that is
\begin{equation}\label{reali}
u(x,t) = U(x,t) \exp\left(i\sum_{j=1}^n \beta_j x_j^2\right),
\end{equation}
where $U(x,t)$ is a \emph{real} function. 

Why use a quadratic phase? Although there is not a formal justification and we do not know of any rigorous error bounds for the method to be presented here, there are several reasons which can heuristically support the use of this ansatz for the phase for situations where there are no essential shape changes of the solutions during the evolution. First of all, when Eq. (\ref{NLSredu}) has self-similar solutions, they have exactly a quadratic phase \cite{PhD}. 
 Secondly, the dynamics of the phase close to 
stationary solutions of the classical cubic NLS equation in two spatial dimensions (critical case) is known to be approximated by quadratic phases \cite{SIAMFibich,Fibich2001}
Finally,  to capture the dynamics of the phase
of solutions close to the stationary ones, which have a constant
phase, by means of a polynomial fit,  the terms of lowest order are quadratic since
the linear terms in the phase may be eliminated by using Theorem \ref{VadVad}.

We would like to remark that a frequently used approach in the physical sciences to study the evolution of solutions of Eq. (\ref{NLS}) is the so called collective coordinate method (also called averaged lagrangian method or variational method) where an specific ansatz for the solution is proposed depending on a finite number ($S$) of free parameters $u(x,t) = \varphi(x,p_1(t),...p_S(t))$. The equations for the evolution of the parameters are then sought from variational arguments \cite{Borisreview,Angel}. For NLS equations all commonly used ansatzs have a 
 quadratic phase. Our systematic method provides a more general framework in which other methods can be systematized and understood.

As we will see in what follows, the choice (\ref{reali}) allows to obtain explicit evolution equations and solves the problem of calculating the integrals of the phase
derivatives in the equations \eqref{momentos-teor}.

\subsection{Modulated power-type nonlinear terms}

Under the QPA, for modulated power-type nonlinearities $g(\rho,t) = g_0(t)
\rho^{p/2}$, $p \in \mathbb{R}$ for which $\int_{\mathbb{R}^n} D(\rho) dx =
-p J/2$, the moment equations \eqref{momentos-teor} are
\begin{subequations}
\label{fase2}
\begin{eqnarray}
\frac{dI_{2,j}}{dt} & = & V_{1,j},\label{ba}\\ \frac{dV_{1,j}}{dt}
& = & 4K_j-2\lambda_j^2I_{2,j}+ p J,\\ \frac{dK_j}{dt} & = &
-\frac{1}{2}\lambda_j^2V_{1,j} + p \beta_j J, \label{bc}
\\ \frac{dJ}{dt} & = & - p \left(\sum_{j=1}^n
\beta_j\right) J + \frac{1}{g_0}\frac{dg_0}{dt} J\label{bd}.
\end{eqnarray}
\end{subequations}
To these equations we must add the identity $V_{1,j} = 4 \beta_j I_{2,j}$,
which is directly obtained by calculating $V_{1,j}$. Or, expressed
otherwise
\begin{equation}
\beta_j = \frac{\dot{I}_{2,j}}{4I_{2,j}}.
\end{equation}

Let us now consider the simplest case of solutions with spherical
symmetry with $\lambda_j = \lambda(t)$, $j=1,\ldots,n$, for which
$\phi(x_1,...,x_n) = \beta(t) \left(x_1^2 + ... + x_n^2\right)$. Using the same notation as in Eqs. \eqref{valoresmedios} the moment equations become
\begin{subequations}
\label{nolinio}
\begin{eqnarray}
\frac{d\mathcal{I}}{dt} & = & \mathcal{V},\\
\frac{d\mathcal{V}}{dt} & = &
4\left(\mathcal{K}+\frac{np}{4}J\right)-2\lambda^2\mathcal{I}, \\
\frac{d\mathcal{K}}{dt} & = & -\frac{1}{2}\lambda^2\mathcal{V} +
np\beta J,\\
\frac{dJ}{dt} & = &  -np\beta J +
\frac{1}{g_0}\frac{d g_0}{d t}J
\end{eqnarray}
\end{subequations}
with $ \mathcal{V} = 4\beta \mathcal{I}.$ Despite the complexity
of the system of equations \eqref{nolinio} it is possible to find two positive invariants
\begin{subequations}
\begin{eqnarray}
\mathcal{Q}_1 & = & 2\mathcal{K}\mathcal{I} -
\mathcal{V}^2/4, \\ \mathcal{Q}_2 & = &
\frac{np}{2g_0}\mathcal{I}^{np/4} J.
\end{eqnarray}
\end{subequations}
The existence of these invariants provides $J$ as a function of
$\mathcal{I}$, which allows us to arrive at an equation for $X =
\mathcal{I}^{1/2}$
\begin{equation}\label{lamejor}
\frac{d^2X}{dt^2} + \lambda^2(t) X = \frac{\mathcal{Q}_1}{X^3} + g_0(t)
\frac{\mathcal{Q}_2}{X^{np/2+1}}.
\end{equation}
Again we obtain a Hill's equation with a singular term.  Note that in the case $n
=3, p = 2$ we have a quartic term in the denominator, which
corresponds with the type of powers that appear in the equations which are obtained in the framework of averaged Lagrangian methods \cite{Borisreview}.

\begin{figure}
\begin{center}
\epsfig{file=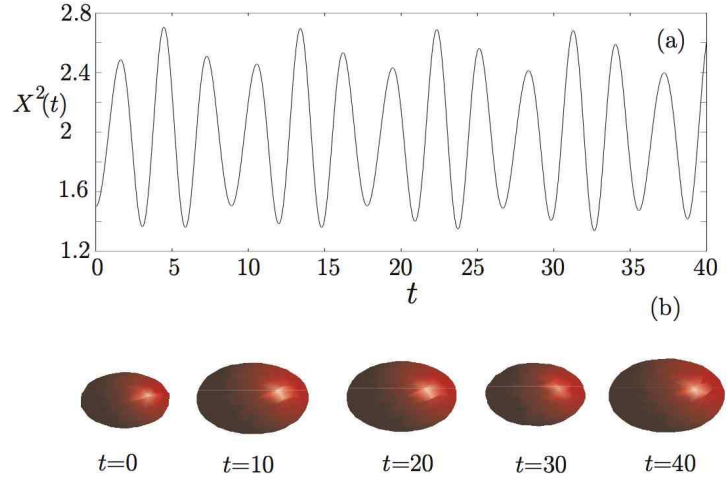,width=0.7\textwidth}
\end{center}\caption{Solutions of \eqref{NLSredu} in three
dimensions with $p=2$, $\lambda^2(t) = 1 + 0.1 \sin (2.8 t)$ and
$g_0 = 10$ with initial data $u_0(x) = e^{-x^2/2}/\pi^{3/4}$. (a)
$X^2(t)$ obtained numerically from the solutions on the 3D grid. (b)
Isosurfaces for $|u|^2 = 0.02$ on the spatial region
$[-3,3]\times[-3,3]\times[-3,3]$ and different instants of time
showing the oscillations of the solution \label{noresonante1}}
\end{figure}

The quadratic phase method provides reasonably precise results
that at least desribe the qualitative behavior of the solutions of the partial differential equation. Using several numerical methods, we
have carried out different tests especially in the most realistic
case $np = 6$ in \eqref{NLSredu}. For example, in figure
\ref{noresonante1} the results of a simulation of \eqref{NLSredu}
with $n=3$, $p=2$, $\lambda^2(t) = 1 + 0.1 \sin (2.8 t)$ and
$g_0=10$ are presented for an initial datum $u_0(x) =
e^{-x^2/2}/\pi^{3/4}$. In this case the simplified equation
\eqref{lamejor} predicts quasi-periodic solutions which is what we
obtain when resolving the complete problem.

\begin{figure}
\begin{center}
\epsfig{file=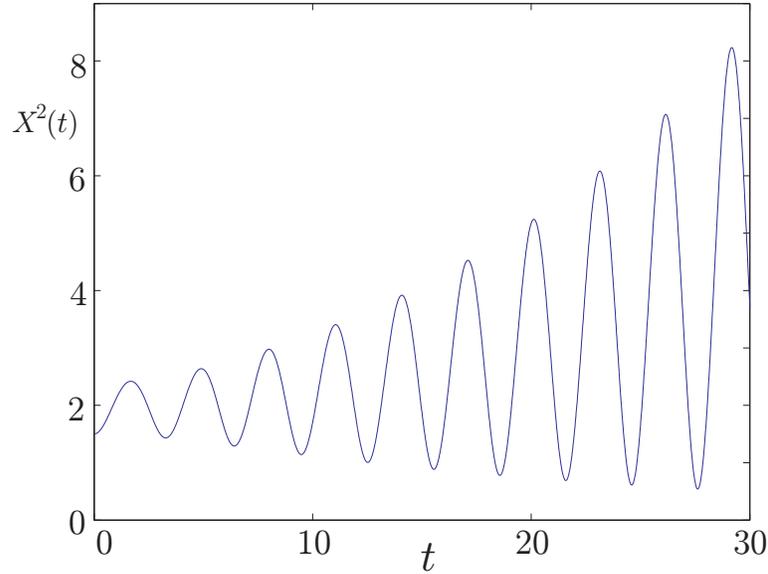,width=0.7\textwidth}
\end{center}\caption{Results of the simulation of \eqref{NLSredu}
in three dimensions on a grid of $64 \times 64 \times 64$ with
$\Delta t = 0.005$ for $p=2$, $\lambda^2(t) = 1 + 0.1 \sin (2.1
t)$. $X^2(t)$ obtained numerically from the solutions on the 3D
grid is shown.\label{resonante1}}
\end{figure}

In Fig. \ref{resonante1}  we show the results for 
$\lambda^2(t) = 1 + 0.1 \sin (2.1 t)$ for which \eqref{lamejor}
predicts resonant solutions. Again, the results of the two models
are in good agreement.

Another interesting application of the quadratic phase
approximation method is the case of cubic nonlinearity, $g(\rho,t) = g_0(t)\rho= g_0(t)|u|^2$, without
potential $\lambda(t)=0$.
In that situation Eq. \eqref{lamejor} becomes
\begin{equation}\label{widthtrapless}
\frac{d^2 X}{dt^2} = \frac{\mathcal{Q}_1}{X^3}+ g_0(t) \frac{\mathcal{Q}_2}{X^{n+1}}
\end{equation}
where the conserved quantities are
\begin{subequations}\label{conservedtrapless}
\begin{eqnarray}
\mathcal{Q}_1 & = & 2\mathcal{K}\mathcal{I}-\mathcal{V}^2/4,\\
\mathcal{Q}_2 & = & \frac{n}{g_0}\mathcal{I}^{n/2}J.
\end{eqnarray}
\end{subequations}
This model describes the propagation of light in nonlinear Kerr
media as well as the dynamics of trapless Bose-Einstein
condensates. In this situation the previous equations are used to
study the possibility of stabilizing unstable solutions of the NLS
equation  by means of an
appropriate temporal modulation of the nonlinear term, that is, by
choosing a suitable function $g_0(t)$ thus providing an alternative to more heuristic treatments \cite{pisaUeda,pisaBoris,Malomed}. More details can be seen in
Ref. \cite{pre2}. 

\subsection{Closure of the equations in other cases}

We have just seen that the quadratic phase approximation method
allows us to close the moment equations in the case of power-type
nonlinear terms. Following those ideas we have managed to close
the equations in more general cases.

We start from the evolution equations for the mean moments
\eqref{cierre}, that after performing the quadratic phase
approximation become
\begin{subequations}\label{momentaaprox}
\begin{eqnarray}
\frac{d\mathcal{I}}{dt} & = & \mathcal{V},\\
\frac{d\mathcal{V}}{dt} & = & 4\mathcal{K}-2\lambda^2\mathcal{I}-2n\int_{\mathbb{R}^n} D(\rho,t), \\
\frac{d\mathcal{K}}{dt} & = & -\frac{1}{2}\lambda^2\mathcal{V}
-2n\beta\int_{\mathbb{R}^n} D(\rho,t),  \\
\frac{dJ}{dt} & = & 2n\beta\int_{\mathbb{R}^n} D(\rho,t)+ \int_{\mathbb{R}^n} \frac{\partial
G(\rho,t)}{\partial t}.
\end{eqnarray}
\end{subequations}
and $\mathcal{V}=4\beta\mathcal{I}$.

The idea to close the previous equations is to calculate
the evolution of $\int_{\mathbb{R}^n} D(\rho,t) dx$, which is the term that prevent us
from closing the equations, and try to write this evolution in
terms of the moments.  Let us define a new moment $\mathcal{F}$ as
\begin{equation}
\mathcal{F}(t) =  \int_{\mathbb{R}^n} D(\rho,t) dx =J-\int_{\mathbb{R}^n} \rho\frac{\partial G(\rho,t)}{\partial\rho} dx.
\label{F}
\end{equation}
Then, the evolution equations are
\begin{subequations}\label{newmomentaeq}
\begin{eqnarray}
\frac{d\mathcal{I}}{dt} & = & \mathcal{V},\\
\frac{d\mathcal{V}}{dt} & = &
4\mathcal{K}-2\lambda^2\mathcal{I}-2n\mathcal{F}, \\
 \frac{d\mathcal{K}}{dt} & = & -\frac{1}{2}\lambda^2\mathcal{V} -2n\beta \mathcal{F},  \\
\frac{dJ}{dt} & = & 2n\beta \mathcal{F}+\int_{\mathbb{R}^n}\frac{\partial G}{\partial t} dx,
\end{eqnarray}
\end{subequations}
together with the evolution of $\mathcal{F}$
\begin{equation}\label{evolF}
\frac{d\mathcal{F}}{dt}=2n\beta F+2n\beta\int_{\mathbb{R}^n} \rho^2\frac{\partial^2
G}{\partial\rho^2}dx+\int_{\mathbb{R}^n}\frac{\partial G}{\partial
t}dx-\int_{\mathbb{R}^n}\rho\frac{\partial}{\partial t}\frac{\partial
G}{\partial\rho}dx.
\end{equation}
To try to close the system of equations
\eqref{newmomentaeq}-\eqref{evolF} we impose that $\int_{\mathbb{R}^n}
\rho^2\frac{\partial^2 G}{\partial\rho^2} dx$ is a linear combination
of $\mathcal{F}$ and $J$
\begin{equation}\label{lincomb}
\int_{\mathbb{R}^n}\rho^2\frac{\partial^2 G}{\partial\rho^2} dx =a_{\mathcal{F}} \mathcal{F}+a_J
J=(a_{\mathcal{F}}+a_J)\int_{\mathbb{R}^n} G dx -a_{\mathcal{F}} \int_{\mathbb{R}^n}\rho\frac{\partial G}{\partial\rho}dx,
\end{equation}
where $a_{\mathcal{F}}$ and $a_J$ are two arbitrary constants. Then, $G$ must verify 
\[
\int_{\mathbb{R}^n} \left[\rho^2\frac{\partial^2 G}{\partial\rho^2}+a_{\mathcal{F}}
\rho\frac{\partial G}{\partial\rho}-(a_{\mathcal{F}}+a_J) G\right]=0.
\]
Therefore, if the nonlinear term $g(\rho)$ in the NLS equation is
such that $G(\rho)$ verifies Euler's equation
\begin{equation}\label{Euler}
\rho^2\frac{\partial^2 G}{\partial\rho^2}+a_{\mathcal{F}} \rho\frac{\partial
G}{\partial\rho}-(a_{\mathcal{F}}+a_J) G=0,
\end{equation}
the evolution equations will close. In that case we can write
$G(\rho,t)=g_0(t)G_1(\rho)$, where $g_0(t)$ is an arbitrary
function which indicates the temporal variation of the nonlinear
term and $G_1(\rho)$ satisfies Eq. \eqref{Euler}. So
\begin{subequations}
\begin{eqnarray*}
\int_{\mathbb{R}^n}\frac{\partial G}{\partial t} dx & = & \frac{dg_0}{dt}\int_{\mathbb{R}^n}
G_1(\rho) dx=\frac{1}{g_0}\frac{dg_0}{dt}J(t),\\
\int_{\mathbb{R}^n}\rho\frac{\partial}{\partial t}\frac{\partial G}{\partial\rho}dx
& = & \frac{dg_0}{dt}\int_{\mathbb{R}^n}\rho\frac{d
G_1}{d\rho}=\frac{1}{g_0}\frac{dg_0}{dt}[J(t)-\mathcal{F}(t)],
\end{eqnarray*}
\end{subequations}
and the moment equations are written as
\begin{subequations}\label{meanmomentaF}
\begin{eqnarray}
\frac{d\mathcal{I}}{dt} & = & \mathcal{V},\\
\frac{d\mathcal{V}}{dt} & = &
4\mathcal{K}-2\lambda^2\mathcal{I}-2n \mathcal{F}, \\
\frac{d\mathcal{K}}{dt} & = & -\frac{1}{2}\lambda^2\mathcal{V}
-2n\beta \mathcal{F},  \\ \frac{dJ}{dt} & = & 2n\beta
\mathcal{F}+\frac{1}{g_0}\frac{dg_0}{dt}J, \\ \frac{d\mathcal{F}}{dt} & = &
2n\beta(1+a_{\mathcal{F}})\mathcal{F}+2n\beta a_J J+\frac{1}{g_0}\frac{dg_0}{dt}\mathcal{F} .
\end{eqnarray}
\end{subequations}

By solving Eq. \eqref{Euler} we obtain specific nonlinear terms
for which the quadratic phase approximation allows us to write closed
equations for the moments. Depending on the parameter $\delta=(1+a_{\mathcal{F}})^2+4 a_J$
there exist three families of solutions
\begin{equation}\label{Eulersol}
G_1(\rho)=\begin{cases} C_1\rho^{p_+}+C_2\rho^{p_-},\ \delta> 0,\\
C_1\rho^R+C_2\rho^R\log \rho,\ \delta=0,\\
 C_1\rho^R\cos(I\log \rho)+C_2\rho^R\sin(I\log \rho),\ \delta<0,
\end{cases}
\end{equation}
where $
p_\pm=\left((1-a_{\mathcal{F}})\pm\delta^{1/2}\right)/2,$ 
$R = \left(1-a_{\mathcal{F}}\right)/2$, $I=|\delta|^{\frac{1}{2}}/2$.

The most interesting case for applications is $\delta>0$ the
nonlinear term being of the form
\begin{equation}\label{g1lincomb}
g_1(\rho)=k_1\rho^{p_+ -1}+k_2\rho^{p_- -1},
\end{equation}
where $k_1$ and $k_2$ are arbitrary constants and $p_+$ and $p_-$ are defined through the relations
\begin{subequations}
\begin{eqnarray}
a_{\mathcal{F}} & = & 1 - p_+ - p_-,\\
 a_J & = & -(p_+ - 1)(p_- - 1).
\end{eqnarray}
\end{subequations}
Eq.
\eqref{g1lincomb} implies that the quadratic phase approximation allows to
close the moment equations for nonlinear terms which can be written as a
linear combination of two arbitrary powers of $|u|$.

As in the previous subsection it is possible to find some invariant
quantities, namely
\begin{subequations}\label{conserved}
\begin{eqnarray}
Q_1 & = & 2\mathcal{K}\mathcal{I}-\mathcal{V}^2/4,\\ Q_+ & = &
C\frac{n}{f_+}{\frac{\mathcal{I}^{a_+ n}}{g_0}}(J+f_+ \mathcal{F}),\\ Q_- &
= & C\frac{n}{f_+}{\frac{\mathcal{I}^{a_- n}}{g_0}}(J+f_- \mathcal{F}),
\end{eqnarray}
where
\end{subequations}
\begin{equation}\label{parameters}
a_\pm  =  \frac{p_\pm-1}{2},\ 
f_\pm =  \frac{1}{p_\mp -1}, \
C  =  \left(1-\frac{f_-}{f_+}\right)^{-1}=\left(1-\frac{p_- -
1}{p_+ - 1}\right)^{-1}.
\end{equation}
These conserved quantities allow us to write a differential
equation for the dynamical width $X(t)=\mathcal{I}^{1/2}$
\begin{equation}\label{width}
\frac{d^2 X}{dt^2}+\lambda^2(t) X = \frac{\mathcal{Q}_1}{X^3}+ g_0(t)
\left(\frac{\mathcal{Q}_-}{X^{2 a_- n+1}}-\frac{\mathcal{Q}_+}{X^{2a_+ n+1}}\right).
\end{equation}

The most interesting kind of nonlinearity in the form of Eq.
(\ref{g1lincomb}) is the so-called cubic-quintic nonlinearity for
which $g_0(t)=1, \  g_1(\rho) = k_1\rho + k_2\rho^2 = k_1|u|^2 +
k_2|u|^4$. Then, we have $p_+ - 1 = 2$, $p_- - 1  =  1$, $a_{\mathcal{F}}  =
-4$, $a_J  =  -2$, $a_+ = 1$, $f_+ =  1$, $a_-  =  1/2$, $f_-  =
1/2$, $C = 2$. The invariant quantities are
\begin{subequations}\label{conservedcq}
\begin{eqnarray}
\mathcal{Q}_1 & = & 2\mathcal{K}\mathcal{I}-\mathcal{V}^2/4,\\
\mathcal{Q}_+ & = & 2 n \mathcal{I}^n (J+\mathcal{F}),\\
\mathcal{Q}_- & = & 2 n \mathcal{I}^{n/2}(J+\mathcal{F}/2),
\end{eqnarray}
\end{subequations}
and the equation for the width is
\begin{equation}\label{widthcq}
\frac{d^2 X}{dt^2}+\lambda^2(t) X = \frac{\mathcal{Q}_1}{X^3}+
\frac{\mathcal{Q}_-}{X^{n+1}}-\frac{\mathcal{Q}_+}{X^{2n+1}}
\end{equation}
These equations contain a finite-dimensional description of the dynamics of localized solutions of the model and are similar to those found under specific assumptions for the profile $u(x,t)$ (see e.g. \cite{cq1,cq2,cq3}). The main difference is that the method of moments allows to obtain the  equations under minimal assumptions on the phase of the solutions and that depend on general integral quantities related to the initial data $\mathcal{Q}_1, \mathcal{Q}_+, \mathcal{Q}_-$. This is an essential advantage over the averaged Lagrangian methods used in the literature for which the specific shape of the solution must be chosen a priori (see also \cite{Borisreview,PhD}).

\section{Approximate methods (II): Thomas-Fermi Limit}
\label{limitetf}

\subsection{Concept}

In the framework of the application of the NLS equations to
Bose-Einstein condensation problems (thus for nonlinearities of
the form $g(\rho)=g_0\rho$),  the Thomas-Fermi limit
corresponds to the case $g_0 \gg 1$ (note that this is only one of
the many different meanings of ``Thomas-Fermi" limit in physics).

Usually, what is pursued in this context is to characterize the ground state, defined as the stationary solution of the NLS equation given by Eq. \eqref{st}
with fixed $L^2$ norm having minimal energy $E$. It is also interesting  to find the dynamics of the
solutions under small perturbations of the ground state solution.

\subsection{Physical treatment}

Let us consider the problem of characterizing the ground state of \eqref{NLSredu}.
The usual ``physical" way of dealing with this problem consists of assuming that if the nonlinear term
is very large then it would be possible to neglect the Laplacian term in
\eqref{NLSredu} (!) and to obtain the ground state solution as
\begin{equation}\label{TF1}
\varphi_{TF}(x) = \sqrt{\left(\frac{\mu - \frac{1}{2}\sum\lambda_j^2
x_j^2} {g_0}\right)_+}.
\end{equation}
The value of $\mu$ is obtained from the condition of normalization
$\norm{\varphi_{TF}}_2 = 1$. This procedure provides a solution
without nodes which is then argued to be an approximation to the
ground state.

This method is used in many applied works, but unfortunately it is
not even self-consistent. Near the zero of the radicand of
\eqref{TF1} the approximation obtained has divergent derivatives,
which contradicts the initial hypothesis of ``smallness" of the Laplacian term. 
Although several numerical results can be obtained using this
approximation, its foundation is very weak.

In order to understand the problem better, we rewrite \eqref{NLSredu}
making the change of variables $\kappa = \mu/g_0$, $\eta =
x/\sqrt{g_0}$, $\psi(\eta) = \varphi\left(x/\sqrt{g_0}\right)$, to
give us the equation
\begin{equation}
-\frac{1}{2} \epsilon^2 \Delta \psi + \frac{1}{2}\left(\sum_j
\lambda_j^2 \eta_j^2 \right) \psi + |\psi|^2 \psi = -\kappa \psi,
\end{equation}
with $\epsilon = 1/g_0$. It is evident that $\epsilon^2 \Delta
\psi$ is a singular perturbation whose effect may not be trivial.

\subsection{The method of moments and the Thomas-Fermi limit}

What can be said for the case of
power-type nonlinearities
in the limit $g \gg 1$ on the basis of the method of moments?
Before making any approximations we  write an evolution equation for $\mathcal{I}$ as follows.
 For the sake of
simplicity, though it is not strictly necessary, we will consider
the case of $\lambda_j = \lambda$ for $j = 1,...,n$ and study
the equations for the mean values \eqref{valoresmedios}. 

First, we write Eqs. \eqref{wa} and \eqref{wb} \begin{subequations}
\label{ThTh}
\begin{eqnarray}
\frac{d\mathcal{I}}{dt} & = & \mathcal{V}, \label{pepaa}\\
\frac{d\mathcal{V}}{dt} & = & 4\left(\mathcal{K}+ \frac{np}{4}
J\right)- 2\lambda^2\mathcal{I} = (4-np) \mathcal{K} + np\mathcal{H}
- \left(2+\frac{np}{2}\right)\lambda^2 \mathcal{I}. \label{pepab}
\end{eqnarray}
\end{subequations}
where $\mathcal{H}$ is the conserved energy.
Combining \eqref{pepaa} and \eqref{pepab} we arrive at
\begin{equation}\label{thom}
\frac{d^2\mathcal{I}}{dt^2} + \left(2+\frac{np}{2}\right)\lambda^2
\mathcal{I} = (4-np) \mathcal{K} + np\mathcal{H}.
\end{equation}
Equation \eqref{thom} is exact. 

The fact that the energy functional $E$
reaches a miminum over $\varphi_0$ implies, by Lyapunov stability,
that initial data $u_0(x) = \varphi_0(x) + \varepsilon \delta(x)$
close to the ground state must remain proximal for sufficiently
small values of $\varepsilon$. 

The only approximation needed to complete our analysis is to assume that when $g\gg 1$ then $J \gg \mathcal{K}$ for the ground state. Notice that this is a much more reasonable assumption than the direct elimination of the second derivative in the evolution equation. Thus, the energy conservation and the previous considerations allow us to affirm that $J(t) \gg \mathcal{K}(t)$ for all times. 

Although these facts can be used to write explicit bounds for $\mathcal{K}$, as a first approximation and just in order to show the power of these ideas we can simply take  $\mathcal{K}\simeq 0$. Under this approximation we have
\begin{equation}\label{thom-app}
\frac{d^2\mathcal{I}}{dt^2} + \left(2+\frac{np}{2}\right)\lambda^2
\mathcal{I}  \approx np\mathcal{H}.
\end{equation}
whose solutions can be obtained explicitly  as
\begin{equation}\label{85}
\mathcal{I}(t) \simeq
\frac{np\mathcal{H}}{\lambda^2\left(2+\frac{np}{2}\right)} + A
\cos \left(\lambda t\sqrt{2+\frac{np}{2}}\right) + B \sen
\left(\lambda t\sqrt{2+\frac{np}{2}}\right).
\end{equation}

The equilibrium point of Eq. \eqref{thom} (corresponding to $A=B=0$) gives us the ``size" of the ground state as a function of the physical parameters. Also the frequency of the oscillations around the
 equilibrium point is immediately obtained from Eq. \eqref{85}
\begin{equation}
\Omega = \lambda \sqrt{2+\frac{np}{2}}.
\end{equation}
We have performed numerical simulations of the partial differential
equations \eqref{NLSredu} to verify this prediction. Specifically,
taking $g = 5000 , 20000$, $\lambda=1$ and initial data of the form $u_0(x) =
\varphi_0((1+\varepsilon)x)/\sqrt{1+\varepsilon}$ for $\varepsilon =
0.01$ and $\varepsilon = 0.02$, we find a numerical frequency of
$\Omega_{\text{num}} = 2.26$ which is in excellent agreement with
the value provided by our Thomas-Fermi formula $\Omega_{\text{TF}} =
\sqrt{8} = 2.24$. 

\section{Summary and Conclusions}

In this paper we have developed the method of moments for Nonlinear Sch\"odinger equations. First we have found the general expressions of the method and classified the nonlinearities for which it allows a closed explicit solution of the evolution of the moments. We have also discussed several applications of the method such as the dynamics of Kerr beams in nonlinear stratified media and the dynamics of parametrically forced Bose-Einstein condensates.

Approximate techniques based on the method of moments are also discussed in this paper. In particular, the quadratic phase approximation is also developed here and applied to different problems, such as the writing of simple equations describing the stabilization of solitonic structures by control of the nonlinear terms and the dynamics of localized structures in cubic-quintic media. Finally, we have also studied the moment equations in the so-called Thomas-Fermi limit.

\bibliographystyle{siam}

\end{document}